\documentclass[11pt]{article}
\usepackage{graphicx}
\usepackage{subfloat}
\usepackage{epsfig}
\usepackage{amsmath}
\usepackage{amsfonts}   
\usepackage{amssymb}

\begin{document}
\date{}
%%%%%%%%%%%%%%%%%%%%
\title{{\bf{\Large Ehrenfest's scheme and thermodynamic geometry in Born-Infeld AdS black holes}}}
\author{
{\bf {{\normalsize Arindam Lala}}$
$\thanks{e-mail: arindam.lala@bose.res.in, arindam.physics1@gmail.com}}
,$~${\bf {{\normalsize Dibakar Roychowdhury}}$
$\thanks{e-mail: dibakar@bose.res.in,  dibakarphys@gmail.com}}\\
 {\normalsize S. N. Bose National Centre for Basic Sciences,}
\\{\normalsize JD Block, Sector III, Salt Lake, Kolkata-700098, India}
\\[0.3cm]}
\maketitle
\begin{abstract}
In this paper we analyze the phase transition phenomena in Born-Infeld AdS black holes using Ehrenfest's scheme of standard thermodynamics. The critical points are marked by the divergences in the heat capacity. In order to investigate the nature of the phase transition, we analytically check both the Ehrenfest's equations near the critical points. Our analysis reveals that this is indeed a second order phase transition. Finally, we analyze the nature of the phase transition using state space geometry approach. This is found to be compatible with the Ehrenfest's scheme.
\end{abstract}
\section{Introduction}
\paragraph{}
Thermodynamics of black holes in Anti de Sitter (AdS) space has received renewed attention since the discovery of the phase transition phenomena in the Schwarzschild AdS background\cite{Page}. Till date several attempts have been made in order to describe the phase transition phenomena in black holes\cite{Davies2}-\cite{Doneva}. All these works are basically based on the fact that at the critical point(s) of phase transition the specific heat of the black hole diverges. Inspite of all these attempts, the issue regarding the classification of the nature of the phase transition in black holes remains highly debatable and worthy of further investigations. At this stage it is worthwhile to mention that in usual thermodynamics it is a general practice to adopt the Ehrenfest's scheme\cite{Zeemansky} in order to classify the phase transition phenomena\cite{Nieu1}-\cite{Jackle}. This is mainly due to two of its basic advantages namely, (i) it is simple and elegant and (ii) it provides a unique way to classify the nature of the phase transition in ordinary thermodynamic systems. Even if a phase transition is not truly a second order, we can determine the degree of its deviation by defining a new parameter called Prigogine-Defay ratio ($ \Pi $)\cite{Nieu1, Nieu2, Jackle}. Inspired by all these facts, we propose a possible way to overcome the long standing problem regarding the classification of phase transition in black holes by incorporating the idea of Ehrenfest's scheme from standard thermodynamics. Since black holes in many respects behave as ordinary thermodynamic objects, therefore the extension of Ehrenfest's scheme to black hole thermodynamics seems to be quite natural. Such an attempt has been triggered very recently\cite{Samanta}-\cite{New1}. 

Constructing gravity theories in presence of various higher derivative corrections to the usual Maxwell action has been a popular topic of research for the past several years\cite{ED1}-\cite{ED8}. Among these non-linear theories of electrodynamics it is the Born-Infeld theory that has earned renewed attention for the past few decades due to its several remarkable features. As a matter of fact Einstein-Born-Infeld theory admits various static black hole solutions that possess several significant qualitative features which is absent in ordinary Einstein-Maxwell gravity. These are given by- (i) Depending on the value of the electric charge ($ Q $) and the Born-Infeld coupling parameter ($ b $) it is observed that a meaningful black hole solution exists only for $ bQ\geq 0.5 $. On the other hand, for $ bQ< 0.5 $ the corresponding extremal limit does not exist. This eventually puts a restriction on the parameter space of BI-AdS black holes\cite{Myung1}. This is indeed an interesting feature which is absent in the usual Einstein-Maxwell theory, (ii) Furthermore, one can note that for $ bQ=0.5 $, which corresponds to the critical Born-Infeld-AdS (BI-AdS) case, there exists a new type of phase transition (HP3) which has identical thermodynamical features as observed during the phase transition phenomena in   non-rotating BTZ black holes\cite{Myung1, Myung2}. This is also an interesting observation that essentially leads to a remarkable thermodynamical analogy which does not hold in the corresponding Reissener-Nordstom-AdS (RN-AdS) limit.

Motivated by all the above mentioned features, in the present work, we therefore aim to carry out a further investigation regarding the thermodynamic behavior of BI-AdS black holes\cite{Myung1},\cite{Cai}-\cite{Tanay} in the framework of standard thermodynamics. We adopt the Ehrenfest's scheme of usual thermodynamics in order to resolve a number of vexing issues regarding the phase transition phenomena in BI-AdS black holes. The critical points correspond to an infinite discontinuity in the specific heat ($ C_{\Phi} $), which indicates the onset of a continuous higher order transition. At this point it is worthwhile to mention that, an attempt to verify the Ehrenfest's equations for charged (RN-AdS) black holes was first initiated in \cite{Rab}. There\cite{Rab}, based on numerical techniques, the authors had computed both the Ehrenfest's equations close to the critical point(s). However, due the presence of infinite divergences in various thermodynamic entities (like, heat capacity etc.), as well as the lack of analytic techniques, at that time it was not possible to check the Ehrenfest's equations exactly at the critical point(s). In order to address the above mentioned issues, the present paper therefore aims to provide an analytic scheme in order to check the Ehrenfest's equations exactly at the critical point(s). Moreover, the present analysis has been generalized taking the particular example of BI-AdS black holes which is basically the non-linear generalization of RN-AdS black holes. Our analysis shows that it is indeed a second order phase transition.

Finally, we apply the widely explored state space geometry approach\cite{Rup2}-\cite{Rup3} to analyze the phase transition phenomena in BI-AdS black holes\cite{Sujoy}-\cite{Rab},\cite{Ferra}-\cite{Sahay1}. Our analysis reveals that the scalar curvature ($ R $) diverges exactly at the critical points where $ C_{\Phi} $ diverges. This signifies the presence of a second order phase transition, thereby vindicating our earlier analysis based on Ehrenfest's scheme. It is also reassuring to note that, in the appropriate limit ($ b \rightarrow \infty $, $ Q\neq 0 $) one recovers corresponding results for RN-AdS black holes\cite{Rab}, where $ b $ is the Born-Infeld parameter and $ Q $ is the charge of the black hole.

Before we proceed further, let us mention about the organization of our paper. In section 2 we have discussed  thermodynamics of the BI black holes in AdS space. Using Ehrenfest's scheme, the nature of the phase transition has been discussed in section 3. In section 4 we analyze the phase transition using the (thermodynamic) state space geometry approach. Finally we draw our conclusion in section 5.
\section{Thermodynamic variables of the BI-AdS black holes}
\paragraph{}
In order to obtain a finite total energy for the field around a point-like charge, Born and Infeld proposed a non-linear electrodynamics\cite{Born} in 1934. The Born-Infeld black hole solution with or without a cosmological constant is a non-linear extension of the Reissner-Nordstr\"{o}m black hole solution. In this paper we will be concerned with the Born-Infeld solution in (3+1)-dimensional AdS space-time (with a negative cosmological constant $ \Lambda =-3/l^{2} $) which is given by\cite{Myung1},
\begin{equation}
ds^2=-f(r)dt^2+\dfrac{1}{f(r)}dr^2 + r^{2}d\Omega^{2}
\end{equation}
where we have taken the gravitational constant $ G=1 $. Here the metric coefficient $ f(r) $ is given by,
\begin{equation}
f(r)=1-\dfrac{2M}{r}+r^{2}+\dfrac{2b^{2}r^{2}}{3}\left( 1-\sqrt{1+\dfrac{Q^2}{b^{2}r^4}}\right)+\dfrac{4Q^2}{3r^2}\mathcal{F}\left(\dfrac{1}{4},\dfrac{1}{2},\dfrac{5}{4},\dfrac{-Q^2}{b^{2}r^4} \right)
\end{equation}
where $ \mathcal{F}\left(\dfrac{1}{4},\dfrac{1}{2},\dfrac{5}{4},\dfrac{-Q^2}{b^{2}r^4} \right) $ is the hypergeometric function\cite{Handbook} and $ b $ is the Born-Infeld parameter. In the limit $ b \rightarrow \infty $, $ Q\neq 0 $ one obtains the corresponding solution for the RN-AdS space-time. At this stage it is reassuring to note that the rest of the analysis has been carried out keeping all terms in the hyper geometric series $\mathcal{F}\left(\dfrac{1}{4},\dfrac{1}{2},\dfrac{5}{4},\dfrac{-Q^2}{b^{2}r^4} \right)$.

The ADM mass of the black hole is defined by $ f(r_{+})=0 $, which yields,
\begin{equation}
M(r_{+},Q,b)=\dfrac{r_{+}}{2}+\dfrac{r_{+}^3}{2}+\dfrac{b^2r_{+}^3}{3}\left(1-\sqrt{1+\dfrac{Q^2}{b^{2}r_{+}^4}}\right)+\dfrac{2Q^2}{3r_{+}}\mathcal{F}\left(\dfrac{1}{4},\dfrac{1}{2},\dfrac{5}{4},\dfrac{-Q^2}{b^{2}r^4} \right)
\end{equation} 
where $ r_{+} $ is the radius of the outer horizon. 
%Before we proceed further, let us mention that in the following analysis we will consider terms only linear in $ (\dfrac{1}{b^{2}}) $ in the series of $ \mathcal{F}\left(1/4,1/2,5/4,-Q^{2}/b^{2}r^{4} \right) $.

Using (2), the Hawking temperature for the BI-AdS black holes may be obtained as\footnote{For details regarding the properties of hyper-geometric function, see \cite{Handbook}.},
\begin{eqnarray}
T&=&\dfrac{1}{4\pi}\left( \dfrac{df(r)}{dr}\right)_{r_{+}}\nonumber\\
     &=&\dfrac{1}{4\pi}\left[\dfrac{1}{r_{+}}+3r_{+}+2b^2r_{+}\left( 1-\sqrt{1+\dfrac{Q^2}{b^{2}r_{+}^4}}\right) \right]
\end{eqnarray}

From the first law of black hole thermodynamics we get, $ dM=TdS+\Phi dQ $. Using this we can obtain the entropy of the black hole as, 
\begin{eqnarray}
S &=& \int_{0}^{r_{+}}\dfrac{1}{T}\left(\dfrac{\partial M}{\partial r} \right)_{Q} dr\nonumber\\
  &=& \pi r_{+}^2
\end{eqnarray}

Substituting (5) in (4) we can rewrite the Hawking temperature as\cite{Myung1},
\begin{equation}
T=\dfrac{1}{4\pi}\left[\sqrt{\dfrac{\pi}{S}}+3\sqrt{\dfrac{S}{\pi}}+\dfrac{2b^2\sqrt{S}}{\sqrt{\pi}}\left( 1-\sqrt{1+\dfrac{Q^2\pi ^2}{b^{2}S^2}}\right) \right] 
\end{equation}
\begin{figure}[h]
\centering
\includegraphics[angle=0,width=15cm,keepaspectratio]{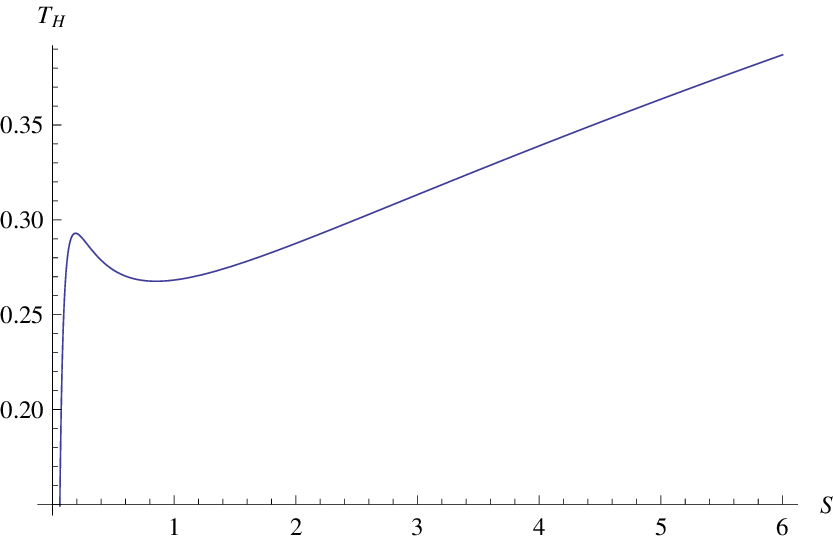}
\caption[]{\it Temperature ($ T $) plot for BI-AdS black hole with respect to entropy ($S$) for fixed $Q=0.13$ and $ b=10 $.}
\label{figure 1}
\end{figure}

We can see from Figure-1 that, there is a `hump' and a `dip' in the $ T-S $ graph. Another interesting thing about the graph is that, it is continuous in $S$. This rules out the possibility of first order phase transition. In order to check whether there is a possibility of higher order phase transition, we compute the specific heat at constant potential ($ C_{\Phi} $) (which is analog of the specific heat at constant pressure ($ C_{P} $) in usual thermodynamics).

The electrostatic potential difference between the black hole horizon and the infinity is defined by\cite{Fer},
\begin{equation}
\Phi =\dfrac{Q}{r_{+}}\mathcal{F}\left(\dfrac{1}{4},\dfrac{1}{2},\dfrac{5}{4},\dfrac{-Q^2}{b^{2}r_{+}^4} \right)
\end{equation}
Using (5) we may further express $ \Phi $ as,
\begin{equation}
\Phi=\dfrac{Q\sqrt{\pi}}{\sqrt{S}}\mathcal{F}\left(\dfrac{1}{4},\dfrac{1}{2},\dfrac{5}{4},\dfrac{-Q^2\pi^{2}}{b^{2}S^{2}} \right)
\end{equation}
From the thermodynamical relation
\begin{center}
$  T=T(S,Q)$
\end{center}
we find,
\begin{equation}
\left(\dfrac{\partial T}{\partial S} \right)_{\Phi}= \left(\dfrac{\partial T}{\partial S} \right)_{Q}-\left(\dfrac{\partial T}{\partial Q} \right)_{S}\left(\dfrac{\partial \Phi}{\partial S} \right)_{Q}\left(\dfrac{\partial Q}{\partial \Phi} \right)_{S}
\end{equation} 

Where we have used the thermodynamic identity
\begin{equation}
\left(\dfrac{\partial Q}{\partial S} \right)_{\Phi}\left(\dfrac{\partial S}{\partial \Phi} \right)_{Q}\left(\dfrac{\partial \Phi}{\partial Q} \right)_{S}=-1.
\end{equation}
Finally using (6), (8) and (9), the heat capacity $ C_{\Phi} $ may be expressed as,
\begin{eqnarray}
C_{\Phi}&=&T\left(\dfrac{\partial S}{\partial T}\right)_{\Phi}\nonumber\\ 
&=&\dfrac{\mathcal{N}(Q,b,S)}{\mathcal{D}(Q,b,S)} 
\end{eqnarray}

where
\begin{eqnarray}
\mathcal{N}(Q,b,S)=-2S\left\lbrace \pi +\left(3-2b^{2}\left(-1+\sqrt{1+\frac{Q^{2}\pi^{2}}{b^{2}S^{2}}}\right)\right)S \right\rbrace\nonumber\\
\left\lbrace b^{2}S^{2}+(\pi^{2}Q^{2}+b^{2}S^{2})\mathcal{F}\left(\dfrac{3}{4},1,\dfrac{5}{4},\dfrac{-Q^2\pi^{2}}{b^{2}S^{2}} \right)\right\rbrace
\end{eqnarray}
 and 
\begin{eqnarray}
\mathcal{D}(Q,b,S)=b^{2}S^{2}\left\lbrace \pi +\left(-3+2b^{2}\left(-1+\sqrt{1+\frac{Q^{2}\pi^{2}}{b^{2}S^{2}}}\right)\right)S \right\rbrace\nonumber\\
+2b^{2}S(-\pi Q+bS)(\pi Q+bS)\mathcal{F}\left(\dfrac{1}{4},\dfrac{1}{2},\dfrac{5}{4},\dfrac{-Q^2\pi^{2}}{b^{2}S^{2}} \right)\nonumber\\
+(\pi -3S-2b^{2}S)(\pi^{2}Q^{2}+b^{2}S^{2})\mathcal{F}\left(\dfrac{3}{4},1,\dfrac{5}{4},\dfrac{-Q^2\pi^{2}}{b^{2}S^{2}} \right)
\end{eqnarray}
\begin{figure}[h]
\begin{minipage}[b]{0.5\linewidth}
\centering
\includegraphics[angle=0,width=15cm,keepaspectratio]{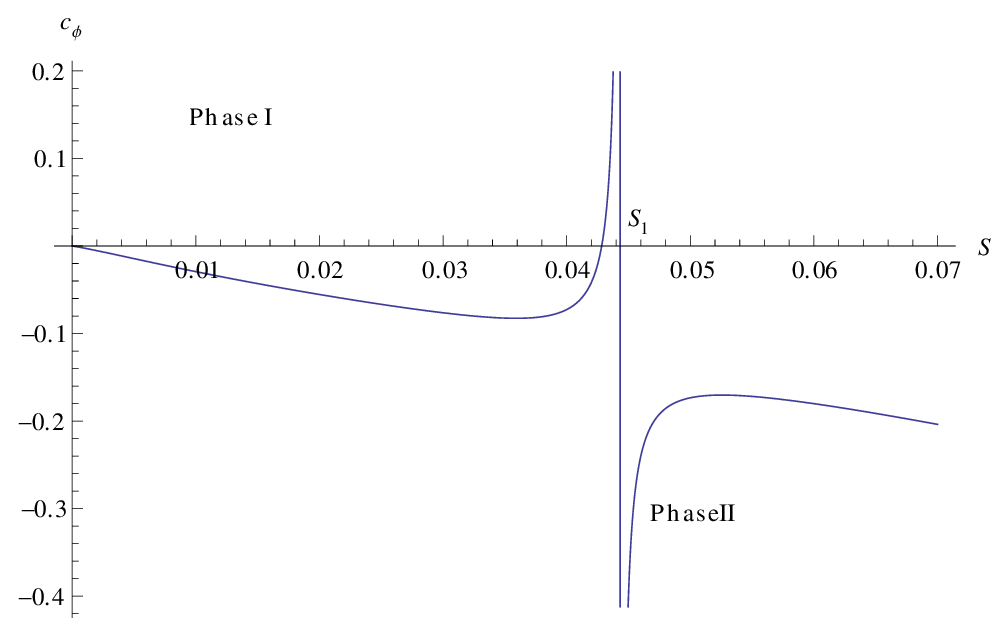}
\caption[]{\it \it Plot of specific heat ($ C_{\Phi} $) against entropy ($S$), at the first critical point ($ S_{1} $), for fixed $Q=0.13$ and $ b=10 $.}
\label{figure 2}
\end{minipage}
\hspace{.1cm}
\begin{minipage}[b]{0.5\linewidth}
\centering
\includegraphics[angle=0,width=15cm,keepaspectratio]{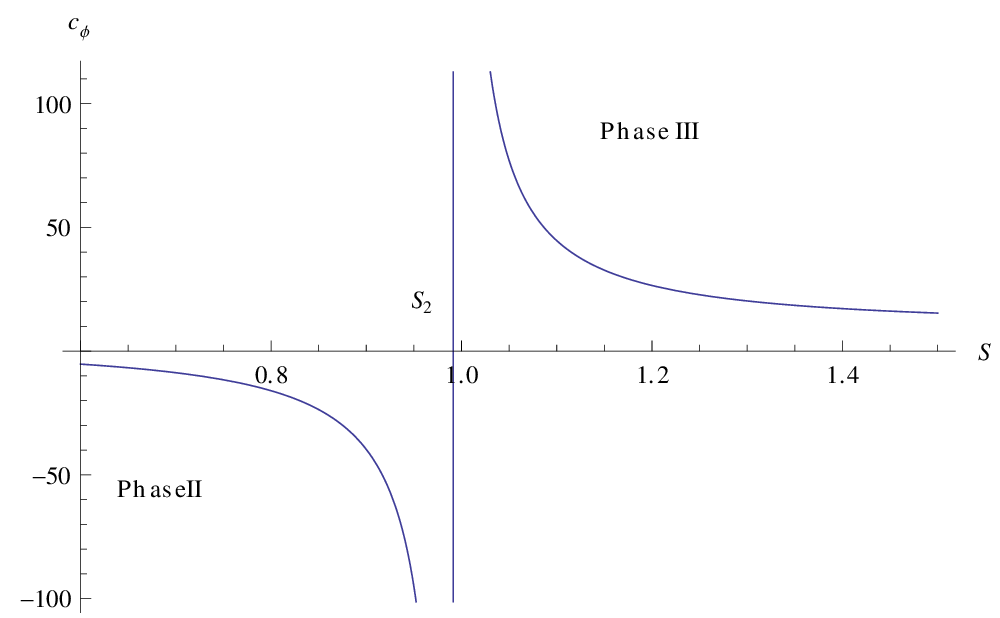}
\caption[]{\it Plot of specific heat ($ C_{\Phi} $) against entropy ($S$), at the second critical point ($ S_{2} $), for fixed $Q=0.13$ and $ b=10 $.}
\label{figure 3}
\end{minipage}
\end{figure}

Before going into the algebric details, let us first plot $ C_{\Phi}-S $ graphs (Figure (2) \& Figure (3)) and make a qualitative analysis of the plots .

From these figures it is evident that the heat capacity ($ C_{\Phi} $) suffers discontinuities exactly at two points ($ S_{1} $ and $ S_{2} $) which correspond to the critical points for the phase transition phenomena in BI-AdS black holes. Similar conclusion also follows from the $ T-S $ plot (Figure (1)), where the `hump' corresponds to $ S_{1} $ and the `dip' corresponds to $ S_{2} $.

The graph of $ C_{\Phi}-S $ shows that there are three phases of the black hole - Phase I ($ 0<S<S_{1} $), Phase II ($ S_{1}<S<S_{2} $) and Phase III ($ S>S_{2} $). Since the higher mass black hole possess larger entropy/horizon radius, therefore at $ S_{1} $ we encounter a phase transition from a  smaller mass black hole (phase I) to an intermediate (higher mass) black hole (phase II). On the other hand, $ S_{2} $ corresponds to the critical point for the phase transition from the intermediate black hole (phase II) to  a larger mass black hole (phase III). Finally, from figures (2) and (3) we also note that the heat capacity ($ C_{\Phi} $) is positive for phase I and phase III, whereas it is negative for phase II. Therefore, phase I and phase III correspond to the thermodynamically stable phases ($ C_{\Phi}>0 $), whereas phase II stands for a thermodynamically unstable phase ($ C_{\Phi}<0 $).

From the above discussions it is evident that the phase transitions we encounter in BI-AdS black holes are indeed continuous higher order. In order to address it more specifically (i.e. whether it is a second order or any higher order transition), we adopt a specific scheme, known as Ehrenfest's scheme in standard thermodynamics.
\section{Study of phase transition using Ehrenfest's scheme}
\paragraph{}
Discontinuity in the heat capacity does not always imply a seecond order transition, rather it suggests a continuous higher order transition in general. Ehrenfest's equations play an important role in order to determine the nature of such higher order transitions for various conventional thermodynamical systems\cite{Nieu1}-\cite{Jackle}. This scheme can be applied in a simple and elegant way in standard thermodynamic systems. The nature of the corresponding phase transition can also be classified by applying this scheme. Moreover, even if a phase transition is not a genuine second order, we can determine the degree of its deviation by calculating the Prigogine-Defay (PD) ratio\cite{Nieu1, Nieu2, Jackle}. Inspired by all these facts, we apply a similar technique to classify the phase transition phenomena in (BI-AdS) black holes and check the validity of Ehrenfest's scheme for black holes.

In conventional thermodynamics, the first and the second Ehrenfest's equations\cite{Zeemansky} are given by,
\begin{eqnarray}
\left(\dfrac{\partial P}{\partial T} \right)_{S}&=&\dfrac{1}{VT}\dfrac{C_{P_{2}}-C_{P_{1}}}{\beta_{2}-\beta_{1}}=\dfrac{\Delta C_{P}}{VT\Delta\beta}\\
 \left(\dfrac{\partial P}{\partial T} \right)_{V}&=&\dfrac{\beta_{2}-\beta_{1}}{{\kappa_{2}-\kappa_{1}}}=\dfrac{\Delta\beta}{\Delta\kappa}
\end{eqnarray}
In case of black hole thermodynamics, pressure ($ P $) is replaced by the negative of the electrostatic potential difference ($ -\Phi $) and volume ($ V $) is replaced by charge ($ Q $) of the black hole.
%\begin{center}
%\begin{tabular}{cc}
%\hline Conventional thermodynamics & BIAdS black holes \\ 
%\hline Pressure ($P$) & Potential ($-\Phi$)  \\ 
%\hline Volume ($V$) & Charge ($Q$) \\ 
%\hline 
%\end{tabular}
%\end{center}

Thus, for the black hole thermodynamics, the two Ehrenfest's equations become\cite{Samanta},
\begin{eqnarray}
-\left(\dfrac{\partial \Phi}{\partial T} \right)_{S}&=&\dfrac{1}{QT}\dfrac{C_{\Phi_{2}}-C_{\Phi_{1}}}{\beta_{2}-\beta_{1}}=\dfrac{\Delta C_{\Phi}}{QT\Delta\beta}\\
-\left(\dfrac{\partial \Phi}{\partial T} \right)_{Q}&=&\dfrac{\beta_{2}-\beta_{1}}{{\kappa_{2}-\kappa_{1}}}=\dfrac{\Delta\beta}{\Delta\kappa}
\end{eqnarray}
respectively. Here, the suffices 1 and 2 denote two distinct phases of the system. In addition to that, $ \beta $ is the \textit{volume expansion coefficient} and $ \kappa $ is the \textit{isothermal compressibility} of the system which are defined as,
\begin{eqnarray}
\beta &=& \dfrac{1}{Q}\left(\dfrac{\partial Q}{\partial T} \right)_\Phi  \\
\kappa &=& \dfrac{1}{Q}\left(\dfrac{\partial Q}{\partial \Phi} \right)_{T}
\end{eqnarray}
Using (6) and (8) and considering the thermodynamic relation,
\begin{center}
$  \left(\dfrac{\partial Q}{\partial T} \right)_{\Phi}=-\left(\dfrac{\partial \Phi}{\partial S} \right)_{Q}\left(\dfrac{\partial Q}{\partial \Phi} \right)_{S}\left(\dfrac{\partial S}{\partial T} \right)_{\Phi} $
\end{center}
we can show that,
\begin{equation}
\beta=\dfrac{-8b^{2}\pi^{\frac{3}{2}}S^{\frac{5}{2}}}{\mathcal{D}(Q,b,S)}
\end{equation}
where the denominator was identified earlier ((13)).

In order to calculate $ \kappa $, we make use of the thermodynamic identity,
\begin{equation}
 \left(\dfrac{\partial Q}{\partial \Phi} \right)_{T} \left(\dfrac{\partial \Phi}{\partial T} \right)_{Q}\left(\dfrac{\partial T}{\partial Q} \right)_{\Phi}=-1.
\end{equation}

Using (10) we obtain from (21)
\begin{equation}
\left(\dfrac{\partial Q}{\partial \Phi} \right)_{T}=\dfrac{\left(\dfrac{\partial T}{\partial S} \right)_{Q}\left(\dfrac{\partial Q}{\partial \Phi} \right)_{S}}{\left(\dfrac{\partial T}{\partial S} \right)_{\Phi}}
\end{equation}
Using (6), (8), (9) and (22) we finally obtain,
\begin{eqnarray}
\kappa=\dfrac{\Psi (Q,b,S)}{\mathcal{D}(Q,b,S)}
\end{eqnarray}

where 
\begin{equation}
\Psi (Q,b,S)=\frac{-2b^{2}S^{\frac{3}{2}}}{Q\pi^{\frac{1}{2}}}\left[2\pi^{2}Q^{2}-\pi S \sqrt{1+\frac{Q^{2}\pi^{2}}{b^{2}S^{2}}}+\left(2b^{2}\left(-1+\sqrt{1+\frac{Q^{2}\pi^{2}}{b^{2}S^{2}}} \right)+3\sqrt{1+\frac{Q^{2}\pi^{2}}{b^{2}S^{2}}}  \right)  \right] 
\end{equation}
and the denominator $ \mathcal{D}(Q,b,S) $ is given by (13).

Note that, the denominators of both $ \beta $ and $ \kappa $ are indeed identical with that of $ C_{\Phi} $. This is manifested in the fact that both $\beta$ and $\kappa$ diverge exactly at the point(s) where $C_{\Phi}$ diverges (Figs. 4-7).

\begin{figure}[h]
\begin{minipage}[b]{0.5\linewidth}
\centering
\includegraphics[angle=0,width=15cm,keepaspectratio]{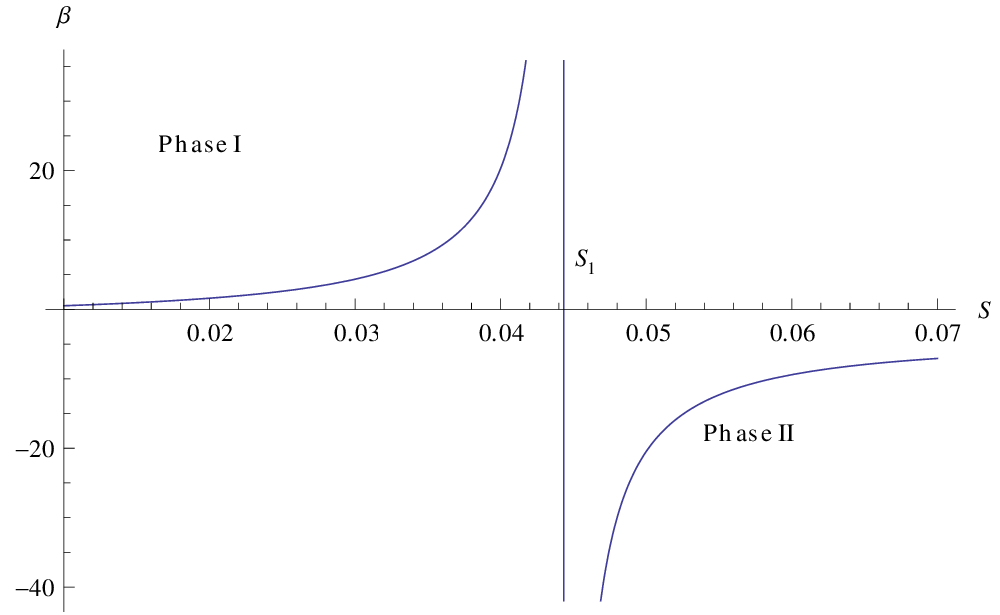}
\caption[]{\it Plot of volume expansion coefficient ($ \beta $) against entropy ($S$), at the first critical point ($ S_{1} $), for fixed $Q=0.13$ and $ b=10 $.}
\label{figure 4}
\end{minipage}
\hspace{.1cm}
\begin{minipage}[b]{0.5\linewidth}
\centering
\includegraphics[angle=0,width=15cm,keepaspectratio]{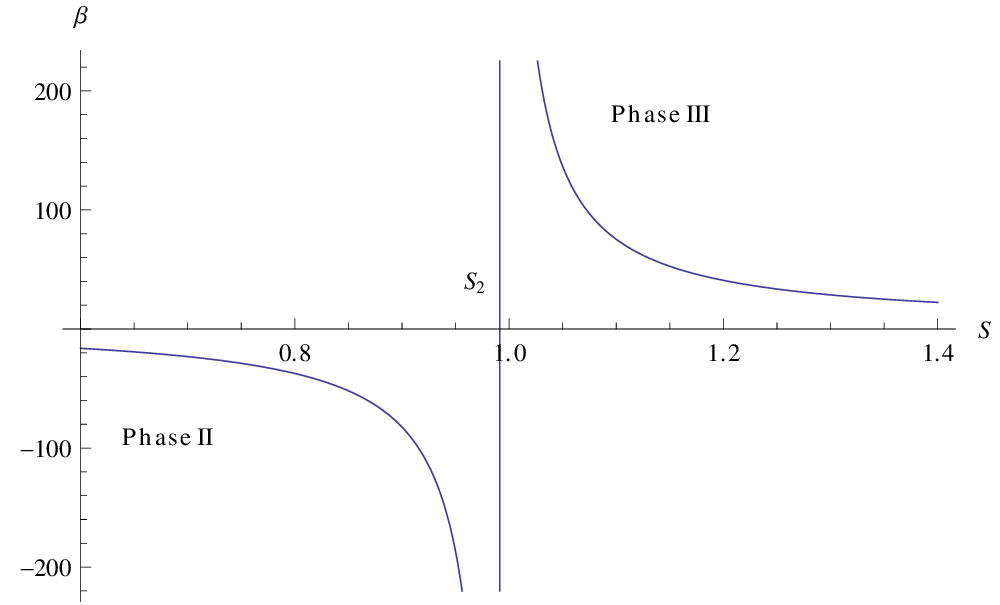}
\caption[]{\it Plot of volume expansion coefficient ($ \beta $) against entropy ($S$), at the second critical point ($ S_{2} $), for fixed $Q=0.13$ and $ b=10 $.}
\label{figure 5}
\end{minipage}
\end{figure}
\begin{figure}[h]
\begin{minipage}[b]{0.5\linewidth}
\centering
\includegraphics[angle=0,width=15cm,keepaspectratio]{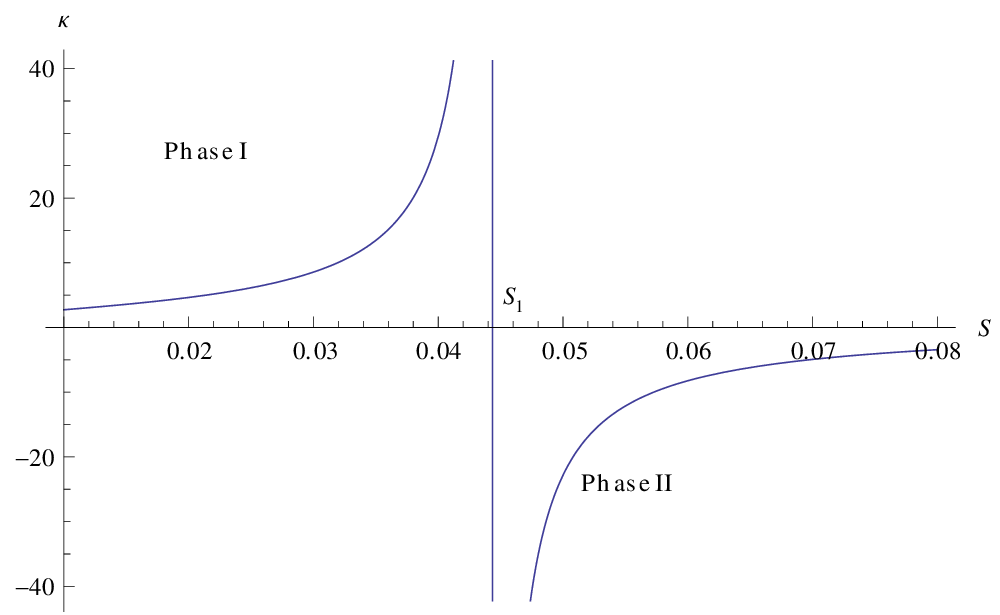}
\caption[]{\it Plot of isothermal compressibility ($ \kappa $) against entropy ($S$), at the first critical point ($ S_{1} $), for fixed $Q=0.13$ and $ b=10 $.}
\label{figure 6}
\end{minipage}
\hspace{.1cm}
\begin{minipage}[b]{0.5\linewidth}
\centering
\includegraphics[angle=0,width=15cm,keepaspectratio]{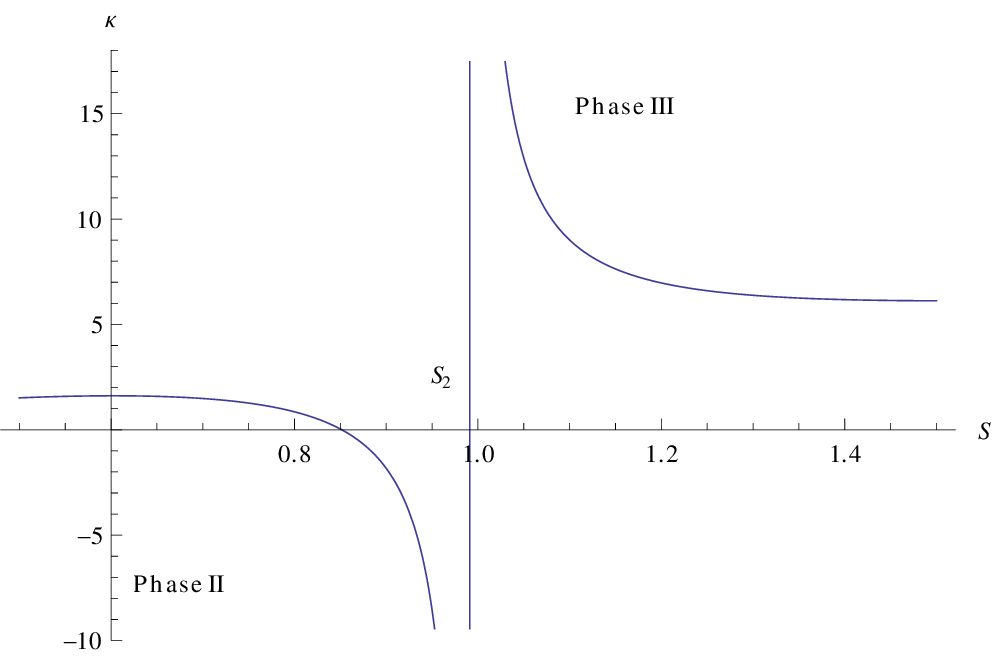}
\caption[]{\it Plot of isothermal compressibility ($ \kappa $) against entropy ($S$), at the second critical point ($ S_{2} $), for fixed $Q=0.13$ and $ b=10 $.}
\label{figure 7}
\end{minipage}
\end{figure}

Being familiar with the two Ehrenfest's equations and knowing $ \beta $ and $ \kappa $, we can now determine the order of the phase transition. In order to do that, we shall analytically check the validity of the two Ehrenfest's equations at the points of discontinuity $ S_{i} $ ( $i=1, 2$). Furthermore, it should be noted that at the points of divergence, we denote the critical values for the temperature ($ T $) and charge ($ Q $) as $ T_{i} $ and $ Q_{i} $ respectively.

Let us now calculate the L.H.S of the first Ehrenfest's equation (16), which may be written as,
\begin{equation}
-\left[\left(\dfrac{\partial \Phi}{\partial T} \right)_{S}\right]_{S=S_{i}} =-\left[\left(\dfrac{\partial \Phi}{\partial Q} \right)_{S}\right]_{S=S_{i}}\left[\left(\dfrac{\partial Q}{\partial T} \right)_{S}\right]_{S=S_{i}}
 \end{equation}
 Using (6) and (8) we further obtain,
 \begin{equation}
-\left[\left(\dfrac{\partial \Phi}{\partial T} \right)_{S}\right]_{S=S_{i}}=\frac{S_{i}}{Q_{i}}\left[1+\left(1+\frac{Q_{i}^{2}\pi^{2}}{b^{2}S_{i}^{2}}\right)\mathcal{F}\left( \frac{3}{4},1,\frac{5}{4},\frac{-Q_{i}^{2}\pi^{2}}{b^{2}S_{i}^{2}} \right)  \right] 
\end{equation}

In order to calculate the R.H.S of the first Ehrenfest's equation (16), we adopt the following procedure. From  (11) and (18) we find,
\begin{eqnarray}
Q_{i}\beta &=&\left[ \left(\dfrac{\partial Q}{\partial T} \right)_{\Phi}\right]_{S=S_{i}}\nonumber\\
            &=&\left[ \left(\dfrac{\partial Q}{\partial S} \right)_{\Phi}\right]_{S=S_{i}}\left(\frac{C_{\Phi}}{T_{i}}\right)  
\end{eqnarray}
Therefore the R.H.S of (16) becomes
\begin{equation}
\dfrac{\Delta C_{\Phi}}{T_{i}Q_{i}\Delta \beta}=\left[ \left(\dfrac{\partial S}{\partial Q} \right)_{\Phi}\right]_{S=S_{i}}
\end{equation}
Using (8) we can further write,
\begin{equation}
\dfrac{\Delta C_{\Phi}}{T_{i}Q_{i}\Delta \beta}=\frac{S_{i}}{Q_{i}}\left[1+\left(1+\frac{Q_{i}^{2}\pi^{2}}{b^{2}S_{i}^{2}}\right)\mathcal{F}\left( \frac{3}{4},1,\frac{5}{4},\frac{-Q_{i}^{2}\pi^{2}}{b^{2}S_{i}^{2}} \right)  \right] 
\end{equation}

From (26) and (29) it is evident that both the L.H.S and the R.H.S of the first Ehrenfest's equation are indeed in good agreement at the critical points $ S_{i} $. As a matter of fact, the divergence of $ C_{\Phi} $  in the numerator is effectively cancelled out by the diverging nature of $ \beta $ appearing at the denominator, which ultimately yields a finite value for the R.H.S of (16).

In order to calculate the L.H.S of the second Ehrenfest's equation, we use the thermodynamic relation, 
\begin{center}
$ T=T(S, \Phi)$
\end{center} 
which leads to 
\begin{equation}
\left(\dfrac{\partial T}{\partial \Phi} \right)_{Q}=\left(\dfrac{\partial T}{\partial S} \right)_{\Phi}\left(\dfrac{\partial S}{\partial \Phi} \right)_{Q}+\left(\dfrac{\partial T}{\partial \Phi} \right)_{S}
\end{equation}
Since $ C_{\Phi} $ diverges at the critical points ($ S_{i} $), it is evident from (11) that 
$ [\left(\frac{\partial T}{\partial S} \right)_{\Phi}]_{S=S_{i}}=0.$ Also from (8) we find that $ \left(\frac{\partial S}{\partial \Phi} \right)_{Q} $ has a finite value at the critical points ($ S_{i} $). Thus from (30) and using (16) we may write,
\begin{equation}
-\left[ \left(\dfrac{\partial \Phi}{\partial T} \right)_{Q}\right]_{S=S_{i}}=-\left[ \left(\dfrac{\partial \Phi}{\partial T} \right)_{S}\right]_{S=S_{i}}=\dfrac{\Delta C_{\Phi}}{T_{i}Q_{i}\Delta \beta}
\end{equation}
From (19), at the critical points we can write,
\begin{equation}
\kappa Q_{i}=\left[ \left(\dfrac{\partial Q}{\partial \Phi} \right)_{T}\right]_{S=S_{i}}
\end{equation}
Using (21) and (18) this can be further written as,
\begin{equation}
\kappa Q_{i}=-\left[ \left(\dfrac{\partial T}{\partial \Phi} \right)_{Q}\right]_{S=S_{i}}Q_{i}\beta
\end{equation}
Therefore finally the R.H.S of (17) may be expressed as\cite{RBDR},
\begin{eqnarray}
\dfrac{\Delta \beta}{\Delta \kappa}&=&-\left[ \left(\dfrac{\partial \Phi}{\partial T} \right)_{Q}\right]_{S=S_{i}} \nonumber\\
&\equiv&-\left[ \left(\dfrac{\partial \Phi}{\partial T} \right)_{S}\right]_{S=S_{i}} \nonumber\\
&=&\frac{S_{i}}{Q_{i}}\left[1+\left(1+\frac{Q_{i}^{2}\pi^{2}}{b^{2}S_{i}^{2}}\right)\mathcal{F}\left( \frac{3}{4},1,\frac{5}{4},\frac{-Q_{i}^{2}\pi^{2}}{b^{2}S_{i}^{2}} \right)  \right] 
\end{eqnarray}
%Using (10) we can further simplify (30) as,
%\begin{equation}
%\dfrac{\Delta \beta}{\Delta \kappa}=\dfrac{2S_{i}}{Q_{i}}\left(1+\dfrac{Q_{i}^{2}\pi^{2}}{5b^{2}S_{i}^{2}} \right)
%\end{equation}
%Once again it is  interesting to note that the divergence of $ \beta $ is cancelled out by that of $ \kappa $ resulting a finite value at the critical points. From (25) and (31) we note that,
%\begin{equation}
%\dfrac{\Delta C_{\Phi}}{T_{i}Q_{i}\Delta \beta}=\dfrac{\Delta \beta}{\Delta \kappa}
%\end{equation}
%Using (27) and (32) we finally observe that,
%\begin{equation}
%-\left[\left( \dfrac{\partial\Phi}{\partial T}\right)_{Q}\right]_{S=S_{i}}=\dfrac{\Delta \beta}{\Delta \kappa}
%\end{equation}
This proves the validity of the second Ehrenfest's equation at the critical points $ S_{i} $. Finally, using (29) and (34) the Prigogine-Defay (PD) ratio\cite{Nieu1} may be obtained as,
\begin{eqnarray}
\Pi &=&\dfrac{\Delta C_{\Phi}\Delta\kappa}{T_{i}Q_{i}(\Delta \beta)^{2}}\nonumber\\
    &=&1
\end{eqnarray}
which confirms the second order nature of the phase transition.
%Considering the transition from phase I to phase II, we can show exactly in the same way that the two Ehrenfest's equations are also satisfied at the point $ S_{1} $.

%Thus we can write the two Ehrenfest's equations in a general form as below:
%\begin{eqnarray}
%-\left(\dfrac{\partial \Phi}{\partial T} \right)_{S_{i}}&=&\dfrac{1}{Q_{ci}T_{ci}}\dfrac{C_{\Phi_{k}}-C_{\Phi_{j}}}{\beta_{k}-\beta_{j}}\nonumber\\
%&=&\dfrac{2S_{i}}{Q_{ci}}\left(1+\dfrac{Q_{ci}^{2}\pi^{2}}{5b^{2}S_{i}^{2}} \right). 
%\end{eqnarray}

%and

%\begin{eqnarray}
%-\left(\dfrac{\partial \Phi}{\partial T} \right)_{Q_{ci}}&=&\dfrac{\beta_{k}-\beta_{j}}{\kappa_{k}-\kappa_{j}}\nonumber\\
%&=&\dfrac{4\pi Q_{ci}\left(1-\frac{Q_{ci}^{2}\pi^{2}}{2b^{2}S_{i}^{2}}\right)}{\left[-1+\frac{3S_{i}}{\pi}+\frac{2b^{2}S_{i}}{\pi}\left(1-\sqrt{1+\frac{Q_{ci}^{2}\pi^{2}}{b^{2}S_{i}^{2}}}\right)+\frac{4\pi Q_{ci}^{2}}{S_{i}\sqrt{1+\frac{Q_{ci}^{2}\pi^{2}}{b^{2}S_{i}^{2}}}}\right]}.
%\end{eqnarray}
%Where, $ k=$ phase II and $ j=$ phase I, for $ i=1 $ and $ k=$ phase III and $ j=$ phase II, for $ i=2 $.

%In the above analysis we have shown that, at the points of divergence of the thermodynamic quantities, the two Ehrenfest's equations hold. This implies that, the phase transition in the BI AdS black hole is a genuine second order phase transition.
\section{Study of phase transition using state space geometry}
\paragraph{}
The validity of the two Ehrenfest's equations near the critical points indeed suggests a second order nature of the phase transition. In order to analyze the phase transition phenomena from a different perspective, we adopt the thermodynamic state space geometry approach\cite{Rup2}-\cite{Rup3}, which has received renewed attention for the past two decades in the context of black hole thermodynamics\cite{Sujoy}-\cite{Rab},\cite{Ferra}-\cite{Sahay1}. This method provides an elegant way to analyze a second order phase transition near the critical point.

In the state space geometry approach one aims to calculate the scalar curvature ($ R $)\cite{Rup1} which suffers a discontinuity at the critical point for the second order phase transition.

%The thermodynamic state space is characterized by the curvature scalar $ R $. In ordinary thermodynamic systems $ |R| $ is found to be proportional to the correlation volume $ \xi^{d} $, where $ \xi $ is the correlation length and $ d $ is the spatial dimentionality of the system. The curvature scalar ($ R $) then provides information about the microscopic behaviour of the system. But in black hole thermodynamics $ R $ is generally associated with the average number of correlated area elements (Planck areas) on the event horizon\cite{Rup1}. Thus $R$ actually determines the fluctuations of these elements. While zero value of the  scalar curvature implies uncorrelated area elements, diverging nature of the scalar curvature indicates the onset of a phase transition leading to highly correlated area elements.

%For any thermodynamic phase transition the higher order nature of the transition is characterized by the divergence property of the specific heat. Interestingly it is found that for a second order phase transition $ R $ also diverges exactly at the same point where the specific heat diverges and changes its sign. This behaviour of $ R $ is not observed for any first order phase transition.

In order to calculate the curvature scalar ($ R $) we need to determine the Ruppeiner metric coefficients which may be defined as\cite{Rup2}-\cite{Rup3},
\begin{equation}
g_{ij}^{R}=-\dfrac{\partial^{2} S(x^{i}) }{\partial x^{i}\partial x^{j}}
\end{equation}
where $ x^{i}=x^{i}(M, Q) $, $ i=1, 2 $, are the extensive variables of the system. From the computational point of view it is convenient to calculate the Weinhold metric coefficients\cite{Wein},
\begin{equation}
g_{ij}^{W}=\dfrac{\partial^{2} M(x^{i}) }{\partial x^{i}\partial x^{j}}
\end{equation}
(where $ x^{i}=x^{i}(S, Q) $, $ i=1, 2 $) that are conformally connected to that of the Ruppeiner geometry through the following map\cite{Ferra, Janyszek}
\begin{equation}
dS_{R}^{2}=\frac{dS_{W}^{2}}{T} 
\end{equation}

In order to calculate $ g_{ij}^{R} $ we choose $ x^{1}=S $ and $ x^{2}=Q $. Finally, using (3), the Ruppeiner metric coefficients may be found as,
\begin{equation}
g_{SS}^{R} = \dfrac{\dfrac{1}{2S}\left[ -1+\dfrac{3S}{\pi}+\dfrac{2b^{2}S}{\pi}\left( 1-\sqrt{1+\dfrac{Q^{2}\pi^{2}}{b^{2}S^{2}}}\right) +\dfrac{4\pi Q^{2}}{S}-\dfrac{2\pi^{3}Q^{4}}{b^{2}S^{3}}\right] }{\left[1+\dfrac{3S}{\pi}+\dfrac{2b^{2}S}{\pi}\left( 1-\sqrt{1+\dfrac{Q^{2}\pi^{2}}{b^{2}S^{2}}}\right) \right] }
\end{equation}
\begin{equation}
g_{SQ}^{R} = \dfrac{\left( \dfrac{-2\pi Q}{S}+\dfrac{\pi^{3}Q^{3}}{b^{2}S^{3}}\right) }{\left[1+\dfrac{3S}{\pi}+\dfrac{2b^{2}S}{\pi}\left( 1-\sqrt{1+\dfrac{Q^{2}\pi^{2}}{b^{2}S^{2}}}\right) \right]}
\end{equation}

and

\begin{equation}
g_{QQ}^{R} = \dfrac{\left(4\pi-\dfrac{6\pi^{3}Q^{2}}{5b^{2}S^{2}} \right) }{\left[1+\dfrac{3S}{\pi}+\dfrac{2b^{2}S}{\pi}\left( 1-\sqrt{1+\dfrac{Q^{2}\pi^{2}}{b^{2}S^{2}}}\right) \right]}
\end{equation}

Using these metric coefficients, the curvature scalar may be computed as,
\begin{equation}
R=\dfrac{\wp(S, Q)}{\Re(S, Q)}
\end{equation}

The numerator $ \wp(S, Q) $ is too much combursome which prevents us to present its detail expression for the present work. However the denominator $ \Re(S,Q) $ may be expressed as,
\begin{eqnarray}
\Re =\sqrt{1+\dfrac{\pi^{2}Q^{2}}{b^{2}S^{2}}}\left[-\pi +\left( -3+2b^{2}\left( -1+\sqrt{1+\dfrac{\pi^{2}Q^{2}}{b^{2}S^{2}}}\right) \right) S \right] \left( \pi^{2}Q^{2}+b^{2}S^{2}\right)\nonumber\\
    (-\pi^{6}Q^{6}+12b^{2}\pi^{4}Q^{4}S^{2}-3b^{2}\pi^{3}Q^{2}S^{3}+b^{2}\pi^{2}Q^{2}\left( 9-2b^{2}\left( 7+3\sqrt{1+\dfrac{\pi^{2}Q^{2}}{b^{2}S^{2}}} \right) \right)\nonumber\\ S^{4}
         +10b^{4}\pi S^{5}+10b^{4}\left( -3+2b^{2}\left( -1+\sqrt{1+\dfrac{\pi^{2}Q^{2}}{b^{2}S^{2}}}\right) \right)S^{6})^{3}
\end{eqnarray}
\begin{figure}[h]
\begin{minipage}[b]{0.5\linewidth}
\centering
\includegraphics[angle=0,width=15cm,keepaspectratio]{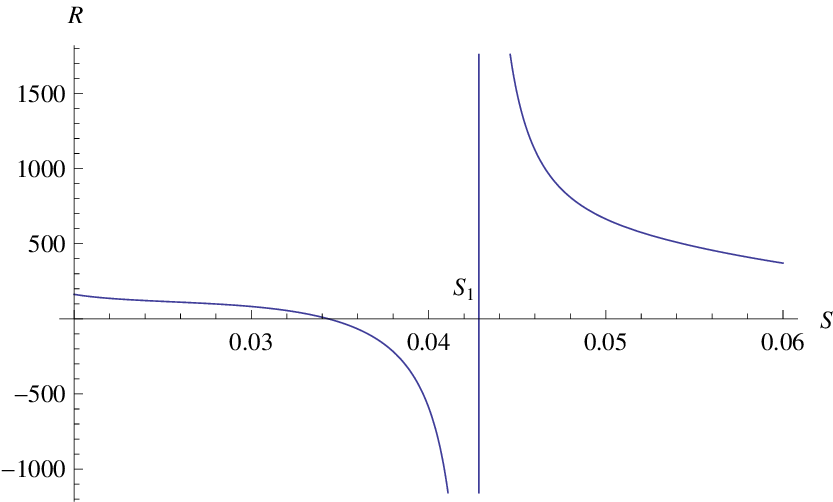}
\caption[]{\it Plot of scalar curvature ($ R $) against entropy ($ S $) at the first point of divergence ($ S_{1} $) (antisymmetric divergence), for fixed $Q=0.13$ and $ b=10 $.}
\label{figure 8}
\end{minipage}
\hspace{.1cm}
\begin{minipage}[b]{0.5\linewidth}
\centering
\includegraphics[angle=0,width=15cm,keepaspectratio]{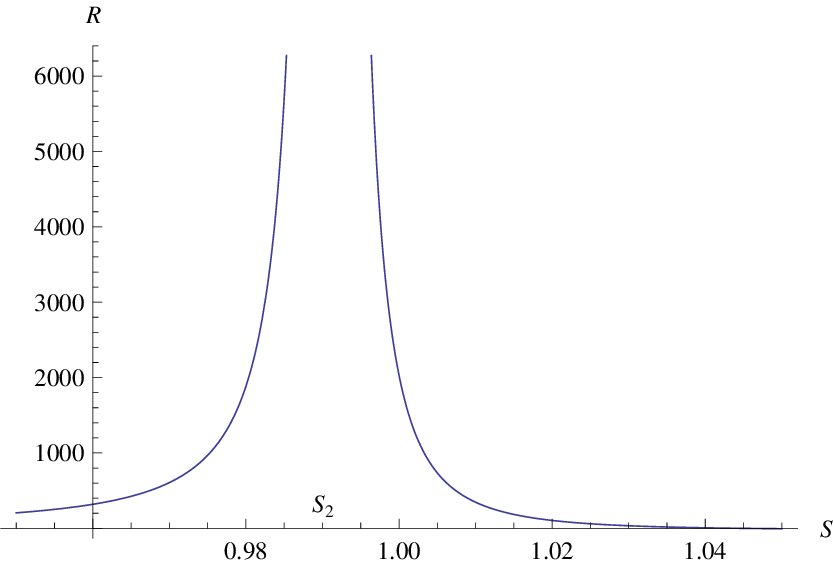}
\caption[]{\it Plot of scalar curvature ($ R $) against entropy ($ S $) at the second point of divergence ($ S_{2} $) (symmetric divergence), for fixed $Q=0.13$ and $ b=10 $.}
\label{figure 9}
\end{minipage}
\end{figure}

Before we conclude this section, let us note few interesting points at this stage. First of all from figures (8) and (9) we observe that the scalar curvature ($R$) diverges exactly at the points where the specific heat ($ C_{\Phi} $) diverges (also see figures (2) \& (3)). This is an expected result, since according to Ruppeiner any divergence in $ R $ implies a corresponding divergence in the heat capacity $ C_{\Phi} $ (which is thermodynamic analog of $ C_{P} $) that essentially leads to a change in stability\cite{Rup1}. On the other hand, no divergence in $ R $ could be observed at the Davis critical point\cite{Davies1}, which is marked by the divergence in $ C_{Q} $ (which is the thermodynamic analog of $ C_{V} $). Similar features have also been observed earlier\cite{Sujoy, Rab}.
 %%%%%%%%%%%%%%%%%%%%%%%%%%%%%%%%%%%%%%%%%%%%%%%%%%%%%%%%%%%%
\section{Conclusions}
\paragraph*{}
In this paper, based on a standard thermodynamic approach, we systematically analyze the phase transition phenomena in Born-Infeld AdS (BI-AdS) black holes. Our results are valid for all orders in the Born-Infeld (BI) parameter ($b$). The continuous nature of the $ T-S $ plot essentially rules out the possibility of any first order transition. On the other hand the discontinuity of the heat capacity ($ C_{\Phi} $) indicates the onset of a continuous higher order transition. In order to address this issue further, we provide a detailed analysis of the phase transition phenomena using Ehrenfest's scheme of standard thermodynamics\cite{Zeemansky} which uniquely determines the second order nature of the phase transition. At this stage it is reassuring to note that the first application of the Ehrenfest's scheme in order to determine the nature of phase transition in charged (Reissener-Nordstrom (RN-AdS)) AdS black holes had been commenced in \cite{Rab}. There the analysis had been carried out \textit{numerically} in order to check the validity of the Ehrenfest's equations close to the critical point(s). Frankly speaking the analysis presented in \cite{Rab} was actually in an underdeveloped stage. This is mainly due to the fact that at that time no such \textit{analytic} scheme was available. As a result, at that stage of analysis it was not possible to check the validity of the Ehrenfest's equations exactly at the critical point(s) due to the occurrence infinite divergences of various thermodynamic entities at the (phase) transition point(s). 

In order to address the above mentioned issue, in the present article we provide an analytical scheme to check the validity of the Ehrenfest's equations exactly at the critical point(s). Furthermore we carry out the entire analysis taking the particular example of BI-AdS black holes which is basically the non-linear generalization of RN-AdS black holes. Therefore our results are quite general and hence is valid for a wider class of charged black holes in the usual Einstein gravity.

Also, we analyze the phase transition phenomena using state space geometry approach. Our analysis shows that the scalar curvature ($ R $) suffers discontinuities exactly at the (critical) points where the heat capacity ($ C_{\Phi} $) diverges. This further indicates the second order nature of the phase transition. Therefore from our analysis it is clear that both the Ehrenfest's scheme and the state space geometry approach essentially lead to an identical conclusion. This also establishes their compatibility while studying phase transitions in black holes.

Finally, we remark that the curvature scalar, which behaves in a very suggestive way for conventional systems, displays similar properties for black holes. Specifically, a diverging curvature that signals the occurence of a second order phase transition in usual systems retains this characteristic for black holes. Our analysis reveals a direct connection between the Ehrenfest's scheme and thermodynamic state space geometry.
\section*{Acknowledgement}
\paragraph*{}
The authors would like to thank the Council of Scientific and Industrial Research (C. S. I. R), Government of India, for financial support. They would also like to thank Prof. Rabin Banerjee for useful discussions.

\end{document}